\documentclass[aip,jcp,amsmath,amssymb,reprint,]{revtex4-1}

\usepackage{dcolumn}
\usepackage{bm}

\usepackage{setspace}

\usepackage{graphicx}
\usepackage{amsfonts,amsmath,amssymb}
\usepackage[caption=false]{subfig}
\usepackage{multirow}

\newcolumntype{d}[1]{D{.}{.}{#1}}

\newcommand{\tabOne}{%
\begin{tabular}{rd{3.2}d{3.2}d{3.2}d{3.2}d{3.2}}
      & \multicolumn{1}{c}{CCSD(T)-F12} & \multicolumn{1}{c}{TTM3-F} & \multicolumn{1}{c}{TTM4-F} & \multicolumn{1}{c}{WHBB5} & \multicolumn{1}{c}{MB-pol} \\
      \firsthline    
2B    & -38.65 & -41.83 & -34.96 & -38.59 & -38.96 \\
3B    &  -8.78 &  -5.02 &  -8.41 &  -9.19 &  -8.74 \\
4B    &  -0.66 &  -0.42 &  -0.48 &  -0.42 &  -0.52 \\
5B    &   0.06 &   0.03 &   0.06 &   0.03 &   0.05 \\
6B    &   0.00 &   0.00 &   0.00 &   0.00 &   0.00 \\
Total & -48.03 & -47.24 & -43.79 & -48.17 & -48.17 \\
\noalign{\smallskip}
\end{tabular}
}

\newcommand{\tabTwo}{%
\begin{tabular}{rd{3.2}d{3.2}d{3.2}d{3.2}d{3.2}}
      & \multicolumn{1}{c}{CCSD(T)-F12} & \multicolumn{1}{c}{TTM3-F} & \multicolumn{1}{c}{TTM4-F} & \multicolumn{1}{c}{WHBB5} & \multicolumn{1}{c}{MB-pol} \\
      \firsthline    
2B    & -38.20 & -42.01 & -36.43 & -37.99 & -38.48 \\
3B    &  -9.04 &  -4.77 &  -8.92 &  -9.45 &  -8.93 \\
4B    &  -0.53 &  -0.26 &  -0.43 &  -0.26 &  -0.47 \\
5B    &   0.01 &   0.01 &   0.03 &   0.01 &   0.03 \\
6B    &   0.00 &   0.00 &   0.00 &   0.00 &   0.00 \\
Total & -47.77 & -47.02 & -45.74 & -47.69 & -47.85 \\
\noalign{\smallskip}
\end{tabular}
}

\newcommand{\tabThree}{%
\begin{tabular}{rd{3.2}d{3.2}d{3.2}d{3.2}d{3.2}}
      & \multicolumn{1}{c}{CCSD(T)-F12} & \multicolumn{1}{c}{TTM3-F} & \multicolumn{1}{c}{TTM4-F} & \multicolumn{1}{c}{WHBB5} & \multicolumn{1}{c}{MB-pol} \\*
\firsthline          
2B    & -35.76 & -40.22 & -35.21 & -35.53 & -35.80 \\
3B    & -10.46 &  -5.65 & -10.05 &  -9.99 & -10.26 \\
4B    &  -1.08 &  -0.56 &  -0.89 &  -0.56 &  -0.92 \\
5B    &  -0.04 &  -0.02 &  -0.01 &  -0.02 &  -0.01 \\
6B    &   0.00 &   0.00 &   0.00 &   0.00 &   0.00 \\
Total & -47.34 & -46.46 & -46.16 & -46.10 & -47.00 \\
\noalign{\smallskip}
\end{tabular}
}

\newcommand{\tabFour}{%
\begin{tabular}{rd{3.2}d{3.2}d{3.2}d{3.2}d{3.2}}
      & \multicolumn{1}{c}{CCSD(T)-F12} & \multicolumn{1}{c}{TTM3-F} & \multicolumn{1}{c}{TTM4-F} & \multicolumn{1}{c}{WHBB5} & \multicolumn{1}{c}{MB-pol} \\*
\firsthline    
2B    & -35.86 & -40.35 & -35.61 & -35.65 & -35.93 \\
3B    & -10.18 &  -5.28 &  -9.75 &  -9.71 & -10.03 \\
4B    &  -1.00 &  -0.49 &  -0.81 &  -0.49 &  -0.85 \\
5B    &  -0.02 &  -0.01 &   0.00 &  -0.01 &   0.00 \\
6B    &   0.00 &   0.00 &   0.00 &   0.00 &   0.00 \\
Total & -47.06 & -46.14 & -46.17 & -45.86 & -46.81 \\
\noalign{\smallskip}
\end{tabular}
}

\newcommand{\tabFive}{%
\begin{tabular}{rd{3.2}d{3.2}d{3.2}d{3.2}d{3.2}}
      & \multicolumn{1}{c}{CCSD(T)-F12} & \multicolumn{1}{c}{TTM3-F} & \multicolumn{1}{c}{TTM4-F} & \multicolumn{1}{c}{WHBB5} & \multicolumn{1}{c}{MB-pol} \\*
      \firsthline    
2B    & -35.01 & -39.71 & -34.99 & -34.89 & -35.24 \\
3B    & -10.43 &  -5.42 &  -9.60 &  -9.86 & -10.15 \\
4B    &  -1.16 &  -0.59 &  -0.86 &  -0.59 &  -0.90 \\
5B    &  -0.02 &  -0.02 &   0.00 &  -0.02 &  -0.01 \\
6B    &   0.01 &   0.00 &   0.00 &   0.00 &   0.00 \\
Total & -46.61 & -45.74 & -45.45 & -45.36 & -46.30 \\
\noalign{\smallskip}
\end{tabular}
}

\newcommand{\tabSix}{%
\begin{tabular}{rd{3.2}d{3.2}d{3.2}d{3.2}d{3.2}}
      & \multicolumn{1}{c}{CCSD(T)-F12} & \multicolumn{1}{c}{TTM3-F} & \multicolumn{1}{c}{TTM4-F} & \multicolumn{1}{c}{WHBB5} & \multicolumn{1}{c}{MB-pol} \\*
\firsthline          
2B    & -32.42 & -37.85 & -32.43 & -32.06 & -32.44 \\
3B    & -11.87 &  -6.66 & -11.13 & -10.71 & -11.64 \\
4B    &  -1.78 &  -1.03 &  -1.43 &  -1.03 &  -1.45 \\
5B    &  -0.19 &  -0.12 &  -0.10 &  -0.12 &  -0.10 \\
6B    &  -0.01 &  -0.01 &   0.00 &  -0.01 &   0.00 \\
Total & -46.27 & -45.66 & -45.09 & -43.92 & -45.63 \\
\noalign{\smallskip}
\end{tabular}
}

\newcommand{\tabSeven}{%
\begin{tabular}{rd{3.2}d{3.2}d{3.2}d{3.2}d{3.2}}
      & \multicolumn{1}{c}{CCSD(T)-F12} & \multicolumn{1}{c}{TTM3-F} & \multicolumn{1}{c}{TTM4-F} & \multicolumn{1}{c}{WHBB5} & \multicolumn{1}{c}{MB-pol} \\*
      \firsthline
2B    & -32.02 & -37.54 & -32.33 & -31.79 & -32.02 \\
3B    & -11.44 &  -6.22 & -10.90 & -10.35 & -11.30 \\
4B    &  -1.63 &  -0.89 &  -1.33 &  -0.89 &  -1.35 \\
5B    &  -0.16 &  -0.09 &  -0.09 &  -0.09 &  -0.09 \\
6B    &  -0.01 &  -0.01 &   0.00 &  -0.01 &   0.00 \\
Total & -45.26 & -44.75 & -44.65 & -43.13 & -44.76 \\
\noalign{\smallskip}
\end{tabular}
}

\newcommand{\tabEight}{%
\begin{tabular}{rd{3.2}d{3.2}d{3.2}d{3.2}d{3.2}}
      & \multicolumn{1}{c}{CCSD(T)-F12} & \multicolumn{1}{c}{TTM3-F} & \multicolumn{1}{c}{TTM4-F} & \multicolumn{1}{c}{WHBB5} & \multicolumn{1}{c}{MB-pol} \\*
      \firsthline    
2B    & -31.96 & -37.55 & -32.45 & -31.66 & -32.02 \\
3B    & -11.43 &  -6.35 & -10.93 & -10.45 & -11.29 \\
4B    &  -1.61 &  -0.90 &  -1.33 &  -0.90 &  -1.35 \\
5B    &  -0.16 &  -0.09 &  -0.09 &  -0.09 &  -0.09 \\
6B    &  -0.01 &  -0.01 &   0.00 &  -0.01 &   0.00 \\
Total & -45.16 & -44.91 & -44.81 & -43.11 & -44.76 \\
\noalign{\smallskip}
\end{tabular}
}

\begin{document}

\title{On the representation of many-body interactions in water}

\author{Gregory R. Medders}
\affiliation{Department of Chemistry and Biochemistry, University of California, San Diego, La Jolla, California 92093, United States}
 
\author{Andreas W. G{\"{o}}tz}
\affiliation{San Diego Supercomputer Center, University of California, San Diego, La Jolla, California 92093, United States}
 
\author{Miguel A. Morales}
\affiliation{Lawrence Livermore National Laboratory, 7000 East Avenue, Livermore, California 94550, U.S.A.}

\author{Pushp Bajaj}
\affiliation{Department of Chemistry and Biochemistry, University of California, San Diego, La Jolla, California 92093, United States}
 
\author{Francesco Paesani}
\affiliation{Department of Chemistry and Biochemistry, University of California, San Diego, La Jolla, California 92093, United States}

\date{\today}

\begin{abstract}

Recent work has shown that the many-body expansion of the interaction energy can be used to develop analytical representations of global potential energy surfaces (PESs) for water. In this study, the role of short- and long-range interactions at different orders is investigated by analyzing water potentials that treat the leading terms of the many-body expansion through implicit (i.e., TTM3-F and TTM4-F PESs) and explicit (i.e., WHBB and MB-pol PESs) representations. It is found that explicit short-range representations of 2-body and 3-body interactions along with a physically correct incorporation of short- and long-range contributions are necessary for an accurate representation of the water interactions from the gas to the condensed phase. Similarly, a complete many-body representation of the dipole moment surface is found to be crucial to reproducing the correct intensities of the infrared spectrum of liquid water. 

\end{abstract}

\maketitle

\section{Introduction}
It is known that the global potential energy surface (PES) of a system containing $N$ interacting water molecules can be formally expressed in terms of the many-body expansion of the interaction energy as a sum over $n$-body terms with $1\le n \le N$,\cite{Hankins_D_1970}
\begin{align}
\label{eq:mb_exp}
  V_N(x_1,\dots,x_N) &= \sum\limits_{a}V^{(1\text{B})}(x_a)
   + \sum\limits_{a>b}V^{(2\text{B})}(x_a,x_b) \nonumber \\
   &+ \sum\limits_{a>b>c}V^{(3\text{B})}(x_a,x_b,x_c)
   + \dots \nonumber\\
   &+ V^{(N\text{B})}(x_1,\dots,x_N).
\end{align}
Here, $x_i$ collectively denotes the coordinates of all atoms in the $i$-th water molecule, $V^{(1\text{B})}$ is the one-body potential describing the energy required to deform an individual water molecule from its equilibrium geometry, and all higher $n$-body interactions, $V^{(n\text{B})}$, are defined recursively through
\begin{align}
  V^{(n\text{B})} &(x_1,\dots,x_n) = V_n(x_1,\dots,x_n)
          - \sum\limits_{a}V^{(1\text{B})}(x_a) \nonumber \\
          &- \sum\limits_{a>b}V^{(2\text{B})}(x_a,x_b)
          - \dots \nonumber \\
          &- \sum\limits_{a_1 > \dots > a_{n-1}}
          V^{((n-1)\text{B})}(x_{a_1},x_{a_2},\dots,x_{a_{n-1}}).
\end{align}
The rapid convergence of Equation (\ref{eq:mb_exp}), demonstrated in several studies of water clusters,\cite{Xantheas_SS_1994,Pedulla_JM_1996,Hodges_MP_1997,Cui_J_2006,Gora_U_2011,Khaliullin_RZ_2012} suggests that the PES associated with a system containing $N$ water molecules can in principle be represented as a sum of low-order interactions that are amenable to accurate calculations using correlated electronic structure methods [e.g., coupled cluster with single, double, and perturbative triple excitations, CCSD(T), which currently represents the gold standard in quantum chemistry]. 

The theoretical modeling of many-body effects in water began in the late 1960s when self-consistent field (SCF) calculations were carried out on small water clusters.\cite{Morokuma_K_1968,Kollman_PA_1969,DelBene_J_1969,DelBene_J_1970,Hankins_D_1970,Hankins_D_1970a,Lentz_BR_1973} These studies concluded that nonadditive effects in water are generally nonnegligible, with 3B contributions being as large as 
10 -- 15\% of the total pair interaction for ring structures. 4B effects were estimated to be, on average, $\sim$1\% of the total pair interaction. The first attempt to derive potential energy functions for water from \textit{ab initio} data was made by Clementi and co-workers who fitted dimer energies calculated at the Hartree-Fock (HF) level of theory to an analytical representation of the 2B interactions.\cite{Popkie_H_1973,Kistenmacher_H_1974,Lie_GC_1976} Later, configuration interaction (CI) calculations were used to derive the pairwise additive MCY potential,\cite{Matsuoka_O_1976} to which 3B and 4B terms were subsequently added through a classical description of polarization effects.\cite{Detrich_J_1984} The first water potential including explicit 2B and 3B terms, derived respectively from 4$^{\text{th}}$ order M\"oller-Plesset (MP4) and HF calculations, along with a classical description of higher-body polarization interactions, was reported in Ref.~\onlinecite{Niesar_U_1989}.  Following these pioneering efforts, the ASP potentials with rigid monomers were derived using intermolecular perturbation theory.\cite{Millot_C_1992,Millot_C_1998}

In the 2000s, Xantheas and co-workers introduced the TTM (Thole-type model) water potentials, which for the first time made use of a highly accurate 1B term derived from \emph{ab initio} calculations.\cite{Burnham_CJ_2002,Xantheas_SS_2002,Burnham_CJ_2002a,Burnham_CJ_2002b} The latest versions (TTM3-F\cite{Fanourgakis_GS_2008} and TTM4-F\cite{Burnham_CJ_2008}) were shown to reproduce the properties of water clusters, liquid water, and ice reasonably well, although some inaccuracies were identified in the calculations of vibrational spectra.\cite{Habershon_S_2008,Paesani_F_2009,Medders2015a}Around the same time, 2B and 3B potentials with rigid water monomers were derived by Szalewicz and co-workers from symmetry-adapted perturbation theory (SAPT).\cite{Mas2000,Mas2003} These studies eventually led to the development of the rigid-monomer CC-pol potential, a 25-site model with explicit 2B and 3B terms fitted to CCSD(T)-corrected MP2 dimer energies and SAPT trimer energies, respectively, with higher-body terms being represented through classical polarization.\cite{Bukowski2007} CC-pol was shown to accurately reproduce the vibration-rotation tunneling spectrum of the water dimer and to predict the structure of liquid water in reasonable agreement with the experimental data. Within the CC-pol scheme, a refined 2B term with explicit dependence on the monomer flexibility has also been developed.\cite{Leforestier_C_2012,Jankowski_P_2015}

More recently, the full-dimensional WHBB\cite{Wang2011a} and MB-pol\cite{Babin2013a,Babin2014,Medders2014} many-body potentials with flexible monomers were developed. WHBB includes explicit 2B and 3B terms fitted to CCSD(T) and MP2 data, respectively, with long-range many-body effects being represented by the same Thole-type model used in the TTM3-F potential. MB-pol was derived from fits to CCSD(T) energies calculated for both water dimers and trimers in the complete basis set (CBS) limit and includes many-body effects within a modified version of the polarization model employed by the TTM4-F potential. Although WHBB has successfully been applied to static calculations of several water properties, to date MB-pol is the only many-body potential that has been consistently employed in quantum molecular dynamics (MD) simulations that accurately predicted the properties of water from the gas to the condensed phase. 

While the calculations with many-body potentials reported in the literature demonstrate that Equation (\ref{eq:mb_exp}) can be effectively used to develop accurate molecular-level representations of the water interactions, the role of short- and long-range contributions at different orders has not been fully investigated. Here, we seek to address this issue by analyzing water potentials that treat the leading terms of the many-body expansion through implicit (e.g., TTM3-F and TTM4-F) and explicit (e.g., WHBB and MB-pol) representations, with a specific focus on how these terms are defined and incorporated as a function of the intermolecular separations. The article is organized as follows: The four potentials are briefly described in section 2, an analysis of the energetics as well as of the vibrational frequencies and intensities for water systems ranging from the gas-phase dimer to the liquid phase is reported in section 3, and the conclusions are given in section 4.

\section{Methods}
The four potentials studied here were chosen because they treat the intramolecular distortions on equal footing, i.e., through the 1B PES developed by Partridge and Schwenke.\cite{Partridge1997a} In this way, the ability of the models to predict water properties is a direct reflection of the treatment of the intermolecular interactions. In addition, WHBB and MB-pol effectively share the same induction scheme employed by TTM3-F and TTM4-F, respectively, which thus allows for a systematic investigation of the effects of explicit low-order terms of the many-body expansion. 

The dominant terms of the many-body water interactions are the 2B and 3B terms, which can be conveniently split into short- and long-range contributions, $V^{(2\text{B},3\text{B})} = V^{(2\text{B},3\text{B})}_{\mathrm{short}} + V^{(2\text{B},3\text{B})}_{\mathrm{long}}$. At the two-body level, long-range interactions are dominated by electrostatics and dispersion, while exchange-repulsion and charge-transfer become increasingly important at short-range. 3B interactions in water, on the other hand, arise primarily from induced interactions at long-range and exchange-repulsion when the monomer electron densities overlap.\cite{Stone1997} Both TTM3-F and TTM4-F include permanent and induced electrostatics as well as dispersion and repulsion at the 2B level, while all higher order terms are represented through many-body induction.\cite{Fanourgakis_GS_2008,Burnham_CJ_2008}

Although WHBB and MB-pol seek to exploit the range separation of the low-order water interactions, the two potentials are intrinsically different both in philosophy and by construction. In WHBB, the short-range 2B term is described by a 7$^\text{th}$-degree permutationally invariant polynomial that smoothly transitions for an oxygen-oxygen separation between 6.5 and 7.5~\AA~into the long-range component described by the 2B TTM3-F potential. The 3B term only includes a short-range component, which is represented by either a 5$^\text{th}$- (WHBB5) or a 6$^\text{th}$-degree (WHBB6) permutationally invariant polynomial that dies off as the largest oxygen-oxygen separation between two molecules of the trimer approaches 8.0 \AA. All higher-order interactions ($n \ge 4$) are described by the polarization model employed in the TTM3-F potential. By construction, WHBB thus employs a strict separation at the 2B and 3B levels between short- and long-range interactions that are described in a completely independent way and are essentially disentangled from all higher-order contributions. It should be noted that a simplified version of WHBB including only 1B, 2B, and 3B interactions with shorter cutoffs was also used,\cite{Liu2015} which, however, cannot be implemented in a straightforward way in standard software for molecular simulations in periodic boundary conditions. For this reason, only calculations with the full WHBB5 implementation of Ref.~\onlinecite{Wang2011a} are reported in the following analysis. 

On the other hand, MB-pol can be viewed as a classical polarizable model supplemented by short-range 2B and 3B terms that effectively represent quantum-mechanical interactions arising from the overlap of the monomer electron densities. Specifically, at all separations, $V^{(2\text{B})}$ includes (damped) dispersion forces derived from \textit{ab initio} computed asymptotic expansions of the dispersion energy along with electrostatic contributions due to the interactions between the molecular permanent and induced moments. At short-range, $V^{(2\text{B})}$ is supplemented by a 4$^\text{th}$-degree permutationally invariant polynomial that smoothly switches to zero as the oxygen-oxygen separation in the dimer approaches 6.5 \AA. Similarly, the MB-pol 3B term, $V^{(3\text{B})}$, includes a 3B polarization term at all separations, which is supplemented by a short-range 4$^\text{th}$-degree permutationally invariant polynomial that effectively corrects for the deficiencies of a purely classical representation of the 3B interactions in regions where the electron densities of the three monomers overlap. This short-range 3B potential is smoothly switched off once the oxygen-oxygen separation between any water molecule and the other two water molecules of a trimer reaches a value of 4.5 \AA. In MB-pol, all induced interactions are described through many-body polarization using a slightly modified version of TTM4-F.\cite{Babin2013a} MB-pol thus contains many-body effects at all monomer separations as well as at all orders, in an explicit way up to the third order and in a mean-field fashion at all higher orders. It should be noted that alternative functions to the permutationally invariant polynomials (e.g., GAP\cite{Bartok2015}) have been suggested and successfully employed in modeling short-range many-body interactions.\cite{Bartok2013} 

\section{Results and Discussion}
The effects of the different representations of the many-body interactions employed by TTM3-F, TTM4-F, WHBB, and MB-pol are investigated through the analysis of the energetics of water systems ranging from the gas-phase dimer to the liquid phase. Figure 1a shows a comparison of the dimer binding energies calculated for the global minimum configuration of each PES. Although all potentials predict binding energies within $\sim$0.2 kcal/mol of the CCSD(T)/CBS reference value (independently of whether the dimer geometry is optimized within each PES or the CCSD(T)/CBS dimer geometry is used in the calculations),\cite{Tschumper2002} the analysis of the many-body water interactions reported in Ref.~\onlinecite{Medders2013} shows that the TTM3-F and TTM4-F interaction energies deviate significantly from the CCSD(T) reference data when dimer configurations different from the corresponding equilibrium geometries are considered. 
Specifically, root mean square deviations (RMSDs) of 0.73 kcal/mol and 0.44 kcal/mol were obtained in Ref.~\onlinecite{Medders2013} for TTM3-F and TTM4-F, respectively, from comparisons with $\sim$1,400 CCSD(T)/aug-cc-pVTZ 2B energies corrected for the basis set superposition error. It was instead shown that both WHBB and MB-pol provide a highly accurate representation of the whole multidimensional dimer PES as demonstrated by RMSDs of 0.15 kcal/mol and 0.05 kcal/mol, respectively, obtained from comparisons with more than 40,000 CCSD(T)/CBS 2B energies.\cite{Babin2013a}

\vspace{0.2cm}
\begin{figure*}[t]
\begin{center}
\includegraphics[width=16.2cm]{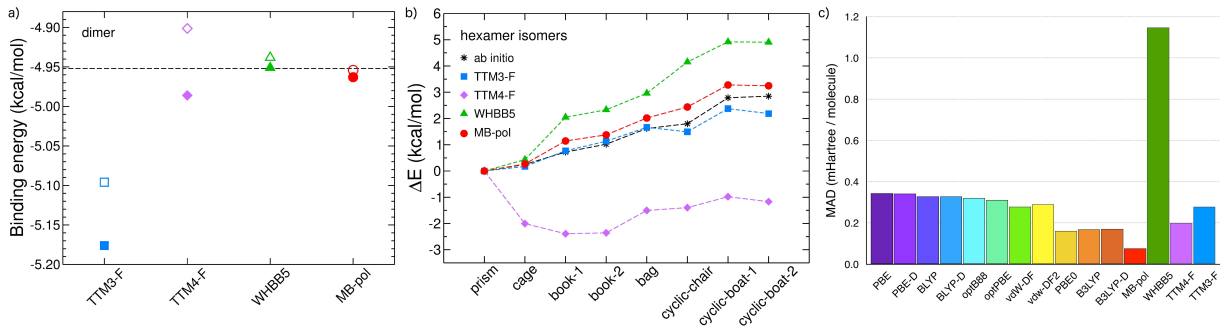}
\caption{a) Binding energies for the global minimum configuration of the water dimer (in kcal/mol) obtained with the TTM3-F, TTM4-F, WHBB, and MB-pol potentials in comparison to the reference CCSD(T) value from Ref.~\onlinecite{Tschumper2002}(filled symbols). Also shown as open symbols are the binding energies calculated using the reference CCSD(T) dimer geometry. b) Comparison of the relative binding energies ($\Delta$E) of the water hexamer isomers calculated with the TTM3-F, TTM4-F, WHBB, and MB-pol potentials using the \emph{ab initio} geometries of Ref.~\onlinecite{Bates2009}. Also shown as a reference are the \emph{ab initio} values from Ref.~\onlinecite{Bates2009}. c) Mean absolute difference in the total energy between QMC and DFT with various exchange-correlation functionals and the TTM3-F, TTM4-F, WHBB, and MB-pol potentials calculated for configurations (in periodic boundary conditions) extracted from path-integral molecular dynamics simulations of water performed with the vdW-DF and vdW-DF2 functionals. The statistical errors in the mean absolute differences arising from the QMC calculations are less than 0.006 mHa/molecule. See Ref.~\onlinecite{Morales2014} for specific details on the QMC and DFT calculations.
}
\end{center}
\end{figure*}

Due to the different treatment of many-body effects, the relative accuracies of the TTM3-F, TTM4-F, WHBB, and MB-pol potentials are expected to differ as the number of molecules in the system of interest increases. This is shown in Figure 1b, where the relative energies of low-lying hexamer isomers calculated with the four potentials are compared with the corresponding \emph{ab initio} values.\cite{Bates2009} The hexamer isomers hold a special place in the experimental and theoretical studies of water because they are the smallest clusters in which the lowest energy configurations correspond to fully three-dimensional structures. For this reason, the hexamer serves as a prototypical model for the hydrogen-bond networks observed in condensed phases. While TTM3-F, WHBB, and MB-pol (qualitatively) reproduce the isomer energy ordering, large deviations are found between the TTM4-F predictions and the reference \emph{ab initio} values. Similar results are also obtained for other small water clusters, including the tetramer and pentamer isomers, shown in Figures S1 and S2 of the supplemental material.\cite{supp_mat}. Since it was shown that the Thole-type polarization scheme employed by the TTM4-F potential correctly captures higher-order water interactions,\cite{Medders2013,Babin2014} the low accuracy shown by TTM4-F in describing the relative energies of the hexamer isomers is possibly due to intrinsic deficiencies in the 2B term. In this context, the performance of the TTM3-F potential, which provides the best agreement with the reference \emph{ab initio} values, is somewhat surprising given the large deviations already seen at both the 2B and 3B levels,\cite{Medders2013} and may suggest some fortuitous cancellation of errors. Based on this, it is interesting to note that the isomer energies predicted by WHBB, which employs the same electrostatic model as TTM3-F, deviate from the reference data by as much as ~2.0 kcal/mol. The origin of these deviations may result from inaccuracies in the short-range 2B and 3B WHBB polynomials, incompatibility between the effective many-body representation encoded in the TTM3-F electrostatic model with the short-range WHBB polynomials, or both. On the other hand, MB-pol, which employs a slightly modified version of the TTM4-F electrostatic model, predicts isomer energies in good agreement with the reference data, suggesting that the MB-pol short-range 2B and 3B terms accurately correct for the deficiencies that are intrinsic to the purely classical representation of these interactions encoded in the TTM4-F electrostatic model.

To provide more quantitative insights into the different performance of the four potentials in describing the relative energies of the hexamer isomers, a many-body decomposition of the corresponding interaction energies (Table~\ref{tab:mbdecomp}) was carried out at the CCSD(T)-F12b/VTZ-F12 level of theory\cite{Knizia2009,Adler2007a} with the VTZ-F12 basis set,\cite{Peterson2008} which was shown to effectively provide the same accuracy as CCSD(T)/CBS calculations at a reduced computational cost.\cite{Tew_DP_2007,Bischoff_FA_2009} As has been well established,\cite{Xantheas_SS_1994,Pedulla_JM_1996,Hodges_MP_1997,Cui_J_2006,Gora_U_2011,Khaliullin_RZ_2012} the interaction energies are dominated by the 2B and 3B terms, with 4B and higher terms making largest (but still relatively small) contributions for cyclic structures.\cite{Xantheas2000} It is interesting to note that, while the hexamer binding energies predicted by TTM3-F are in excellent agreement with the reference calculations (Figure 1b), the results of Table~\ref{tab:mbdecomp} demonstrate that this agreement is clearly achieved by a fortuitous cancellation of errors. For example, in the prism isomer, the TTM3-F 2B interaction is too attractive by $\sim$3.2 kcal/mol while the 3B interaction is too repulsive by $\sim$3.8 kcal/mol, resulting in a total interaction energy that differs from the reference value by merely 0.79 kcal/mol. In contrast, TTM4-F, which accurately represents the electrostatic interactions (as shown by 3B contributions that lie within 1 kcal/mol of the reference values for all low-lying hexamer isomers), is unable to correctly describe the 2B interactions, which have significant contributions from exchange/repulsion and charge transfer. Thus, without the benefit of fortuitous cancellation of errors (that can, to some extent, be encoded into models through empirical parametrization), potentials with accurate electrostatics but without a comparably accurate treatment of short-range quantum mechanical effects may be expected to have relatively poor performance. Having been explicitly derived from many-body expansions of the interaction energies obtained from correlated electronic structure data, both WHBB and MB-pol closely reproduce the CCSD(T)-F12 reference many-body terms. However, nonnegligible discrepancies in the 3B and 4B terms result in WHBB being appreciably less accurate than MB-pol at reproducing the hexamer interaction energies.

\begin{table*}[t]%
\singlespacing
\caption{Many-body decomposition of the interaction energy of low-lying hexamer isomers. All energies are in kcal/mol. The CCSD(T)-F12b\cite{Knizia2009,Adler2007a} calculations were performed with the VTZ-F12 basis set\cite{Peterson2008} and corrected for basis set superposition error through the site-site function counterpoise method\cite{Wells1983} using the Molpro quantum chemistry package.\cite{MOLPRO_brief}}%
\label{tab:mbdecomp}
\centering
\subfloat[][Prism]{\scalebox{0.7}{\tabOne}}%
\qquad
\subfloat[][Cage]{\scalebox{0.7}{\tabTwo}}%

\subfloat[][Book-1]{\scalebox{0.7}{\tabThree}}%
\qquad
\subfloat[][Book-2]{\scalebox{0.7}{\tabFour}}%

\subfloat[][Bag]{\scalebox{0.7}{\tabFive}}%
\qquad
\subfloat[][Cyclic-chair]{\scalebox{0.7}{\tabSix}}%

\subfloat[][Cyclic-boat-1]{\scalebox{0.7}{\tabSeven}}%
\qquad
\subfloat[][Cyclic-boat-2]{\scalebox{0.7}{\tabEight}}%
\end{table*}

The differences between the four potentials become more evident when their accuracy is assessed against benchmark quantum Monte Carlo (QMC) interaction energies calculated for liquid water configurations extracted from path-integral molecular dynamics simulations of the vdW-DF and vdW-DF2 functionals.\cite{Morales2014}
QMC has been shown to be a reliable benchmark in the
study of small water clusters, producing relative energies with an accuracy
comparable to that of CCSD(T).\cite{Santra2008,Gillan2012a}
As a measure of accuracy, the mean absolute deviation (MAD) between the energies $E_i$ obtained with each of the four potentials and the reference QMC energies $E^{\rm QMC}$ was calculated as
\begin{equation}
  {\rm MAD} = \frac{1}{N_c}\sum_{i=1}^{N_c}
  \left| \left(E_i-E_i^{\rm QMC}\right)
  - \left\langle E-E^{\rm QMC}\right\rangle \right|.
\end{equation}
Here, $N_c$ is the number of water configurations and
$\left\langle E-E^{\rm QMC}\right\rangle=(1/N_c)\sum_{i=1}^{N_c}(E_i-E_i^{\rm QMC})$ is the average energy difference for all configurations, effectively aligning the zero of energy with the reference QMC data.
Figure 1c shows that the mean absolute deviation associated with WHBB is $\sim$4 times larger than the corresponding values obtained with both TTM3-F and TTM4-F, and $\sim$15 times larger than the MB-pol value. Interestingly, MB-pol also achieves better accuracy than all DFT models analyzed in Ref.~\onlinecite{Morales2014}. The TTM3-F and TTM4-F results indicate that effective representations of the many-body interactions through classical polarization models can provide a reasonable description of the energetics associated with the liquid phase (i.e., comparable with that provided by most of the DFT models currently used in water simulations), albeit through empirical parameterization. It should be noted that, as shown by the results reported in Table~\ref{tab:mbdecomp} and the many-body analysis of different DFT models carried out in Ref.~\onlinecite{Alfe_D_2014}, cancellation of errors between different terms of the many-body expansion of the interaction energies may also affect the energetics of bulk systems described by both DFT and classical polarizable models. On the other hand, the increased accuracy of MB-pol relative to TTM4-F indicates that short-range many-body interactions beyond a purely classical electrostatic representation are necessary for a correct, molecular-level description of the liquid phase. The significantly different accuracies associated with the WHBB/TTM3-F and MB-pol/TTM4-F potentials also suggest that the correct integration of explicit short-range and effective long-range many-body interactions is critical for ensuring the transferability of the potential across different phases. 

\begin{figure*}[t]
\begin{center}
\includegraphics[width=16.2cm]{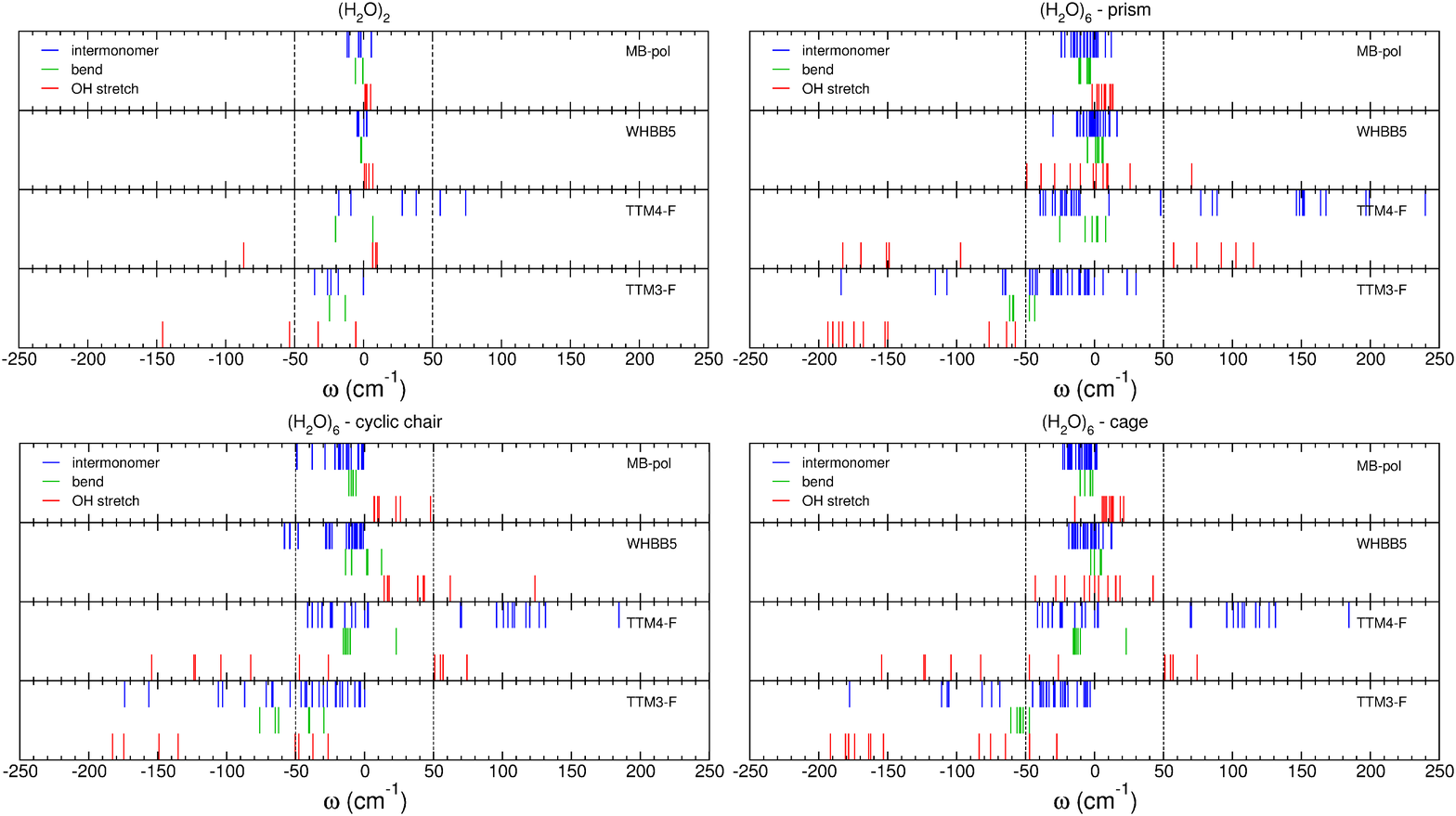}
\caption{Deviations from the 2-body:many-body CCSD(T):MP2 harmonic vibrational frequencies\cite{coleman2015} of (H$_2$O)$_n$ isomers with n = 2 and n = 6 calculated using TTM3-F, TTM4-F, WHBB5, and MB-pol.
}
\end{center}
\end{figure*}

The ability of TTM3-F, TTM4-F, WHBB, and MB-pol to represent the global multidimensional PESs of water systems with an increasing number of molecules is directly reflected in the accuracy with which the four potentials predict the associated vibrational frequencies. Harmonic frequencies for the water hexamer have been reported at the CCSD(T)/aug-cc-pVDZ level.\cite{Miliordos2013} More recently, \emph{ab initio} reference harmonic frequencies for small water clusters have been obtained through two-body:many-body CCSD(T):MP2 calculations using the cc-pVQZ and aug-cc-pVQZ basis sets for H and O atoms, respectively, which enables a more complete treatment of the basis set through the hybrid two-body:many-body approach.\cite{coleman2015} As shown in Figure 2, the WHBB and MB-pol PESs result in harmonic frequencies for the water dimer that deviate by less than 10 cm$^{-1}$ from the \emph{ab initio} frequencies reported in Ref.~\onlinecite{coleman2015}. These results are consistent with the analysis of the dimer vibration-rotation tunneling spectra of WHBB and MB-pol, both of which exhibit nearly perfect agreement with the corresponding experimental data.\cite{Babin2013a} In contrast, relatively large deviations are seen for vibrational frequencies of both the TTM3-F and TTM4-F potentials. The different performance of the four potentials in predicting the relative energies of the hexamer isomers clearly leads to different levels of agreement with the \emph{ab initio} harmonic frequencies. Independently of the isomer, the MB-pol values consistently lie within 50 cm$^{-1}$ of the reference values while the WHBB harmonic frequencies can deviate, in some cases, by more than 100 cm$^{-1}$. Substantially larger deviations, up to $\sim$200 cm$^{-1}$, are instead obtained with both TTM3-F and TTM4-F, reinforcing the notion that purely classical representations of the many-body water interactions are likely not sufficient to fully reproduce the complexity of the underlying multidimensional PESs. Interestingly, all potentials predict somewhat larger deviations for the cyclic-chair isomer of the water hexamer, supporting early observations that many-body effects in water are relatively more important for ring structures.\cite{Lentz_BR_1973}

\begin{figure}[t]
\begin{center}
\includegraphics[width=0.9\columnwidth]{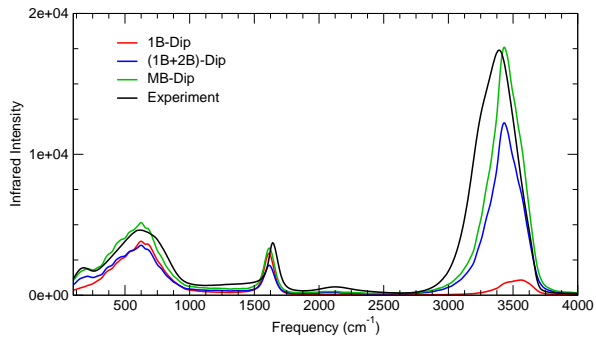}
\end{center}
\caption{\label{fig:mb_elect}Decomposition of the IR spectrum obtained from CMD trajectories with the MB-pol PES in terms of the many-body components of the MB-$\mu$ DMS. 1B-Dip indicates that the one-body (gas-phase monomer) dipoles were used to calculate the dipole of the molecules sampled along the MB-pol CMD trajectories, from which the IR spectrum was calculated. (1B+2B)-Dip indicates that short-ranged two-body dipoles were used in addition to the one-body dipoles. MB-Dip is the full MB-$\mu$ many-body dipole. The spectra were smoothed to facilitate the comparison between the line shapes obtained using the different approximations. The experimental IR spectrum was taken from Ref.~\onlinecite{Bertie1996}. 
}
\end{figure}

While the differences in the PESs clearly affect the underlying vibrational structure, as shown in Figure 2, the infrared activity of those vibrations is ultimately dictated by the associated dipole moment surfaces (DMSs). Many-body representations of higher-order electric properties for molecular systems were characterized beginning in the early 1980s.\cite{Perez1984} The first analysis of many-body effects on the dipole moment of polar molecules was reported by Skwara et al.\cite{Skwara2007} As noted in 1996 by Schwenke,\cite{Schwenke1996} molecular dipole moments can efficiently be represented in terms of geometry-dependent effective charges multiplied by their Cartesian positions. In line with these ideas, the first representation of the 2B dipole of water was reported as part of the WHBB suite.\cite{Wang2011a} A refined version of this 1B + 2B DMS\cite{Liu2015} has been used to calculate the transition dipole moments which were then used to modulate the WHBB5 frequency distributions calculated within the local monomer approximation from configurations of liquid water extracted from MD simulations with the rigid E3B.\cite{Tainter2011} Neglecting both 3B and higher-order contributions to the dipole moments and dynamical effects (e.g., motional narrowing), good agreement in the relative intensities was obtained with the measured IR spectrum of liquid water.\cite{Liu2015} Recently, a complete many-body representation of the DMS of water (MB-$\mu$) was reported,\cite{Medders2015} consisting of the one-body DMS of Lodi et al.\cite{Lodi2011}, an explicit two-body term, and a slightly modified version of TTM4-F polarization for long-range two-body and all higher-order many-body induced dipole moments. The particular functional form of MB-$\mu$ was derived from a systematic analysis of the many-body convergence of the electrostatic properties of water.\cite{Medders2013} 

\begin{table*}[t]
\caption{Cost per molecular dynamics step for different PESs relative to q-TIP4P/F (a non-polarizable model with 3 point charges per molecule and Lennard-Jones interactions between oxygen atoms). The system examined contains 256 water molecules in periodic boundary conditions. Timings are presented for TTM3-F and TTM4-F, the underlying electrostatics model used by WHBB and MB-pol, respectively. The additional cost of WHBB and MB-pol beyond their baseline electrostatics represents the computational cost of the short-range 2B/3B polynomials. WHBB is implemented as described in Ref.~\onlinecite{Wang2011b} using the WHBB polynomial libraries provided by the authors. All timings were performed in a modified version of DL\_POLY2 using a single core of a typical Intel Xeon E5-2640 based workstation.}
\label{tab:perform}
\begin{tabular}{ c c c }
\noalign{\smallskip}
\hline
\hline
    Model & Description & Cost per MD step relative to q-TIP4P/F \\ 
\hline
\noalign{\smallskip}
    q-TIP4P/F & Point charge & 1.0x \\ 
    \noalign{\smallskip}
    TTM3-F & 1 polarizable site/molecule & 7.3x \\ 
    \noalign{\smallskip}
    TTM4-F & 3 polarizable sites/molecule & 8.5x \\ 
    \noalign{\smallskip}
    WHBB & TTM3-F electrostatics + & 29,000x \\ 
         & empirical 2B dispersion + &       \\
         & 2B short-range 7$^{th}$-degree polynomial + &    \\
         & 3B short-range 5$^{th}$-degree polynomial & \\ 
     \noalign{\smallskip}
     MB-pol & TTM4-F electrostatics + & 47x \\ 
         & \emph{ab initio} 2B dispersion + &       \\
         & 2B short-range 4$^{th}$-degree polynomial + &    \\
         & 3B short-range 4$^{th}$-degree polynomial & \\  
\noalign{\smallskip}         
\hline
\hline
\vspace{0.1cm}
\end{tabular}
\end{table*}

To investigate the effects of many-body dipole moments on the IR spectrum of liquid water, many-body molecular dynamics (MB-MD) simulations,\cite{Medders2015} within the centroid molecular dynamics (CMD) formalism, were carried out with the MB-pol PES in combination with the MB-$\mu$ DMS. As shown in Figure 3, while the overall shape of the IR spectrum is captured in the 1B + 2B representation of the DMS, 3B and higher-order contributions to the dipole moment significantly affect the IR intensities. 2B contributions to the DMS lower the bend intensity while they contribute to $\sim$70\% of the intensity of the OH stretch band. Importantly, while 2B contributions are critical to appearance of the shoulder at $\sim$180 cm$^{-1}$ corresponding to the hydrogen-bonding stretch, the 1B contributions are non-negligible. This indicates that some rotational motion is also involved in the hydrogen-bonding stretch. Nevertheless, the absolute intensities of the librational, bending, and stretching bands are only recovered when all many-body effects are included in the calculation of the dipole moment.\cite{Medders2015a} These results support early observations based on molecular dynamics simulations with the SPC potential supplemented with a 3B dipole-induced dipole term showing that the calculated far infrared spectrum of liquid water was in better agreement with experiment relative to results obtained including only a 2B description of the dipole moment.\cite{Guillot_B_1991}

While using a highly accurate PES is a prerequisite for a physically correct description of the water properties at the molecular level, an appropriate balance between accuracy and efficiency is often critical when deciding which potential to employ in computer simulations since the computational cost directly affects the ability to calculate statistically converged quantities. As shown in Table~\ref{tab:perform}, a performance analysis carried out on a single Intel Xeon E5-2640 processor for a system consisting of 256 water molecules in periodic boundary conditions indicates that MB-pol is able to achieve a high level of accuracy at a cost of 47x that of the fixed point charge q-TIP4P/F model\cite{Habershon2009} and $\sim$5.5x that of the TTM4-F potential. WHBB, on the other hand, is $\sim$29,000 more expensive than q-TIP4P/F. A reference implementation of MB-pol is available as an independent plugin\cite{github} for the OpenMM toolkit for molecular simulations.\cite{Eastman_P_2010}

\section{Conclusion}
The role of short- and long-range contributions to the total energy of water systems ranging from the gas-phase dimer to the liquid phase was investigated by considering four water potentials that treat the leading terms of the many-body expansion through implicit (i.e., TTM3-F and TTM4-F PESs) and explicit (i.e., WHBB and MB-pol PESs) representations. The analysis of the energetics, vibrational frequencies, and infrared intensity indicates that explicit short-range representations of 2B and 3B interactions along with a physically correct incorporation of short- and long-range contributions are necessary for an accurate representation of the water interactions, independently of the system size. These results thus suggest that atomistic water potentials built upon the many-body expansion of the interaction energy derived from "first principles" hold great promise to achieve the long-sought-after goal of describing the macroscopic properties of water across different phases from a rigorous microscopic viewpoint. A question that remains to be addressed is the extent to which these "first principles" many-body water potentials can be extended to the modeling of complex aqueous solutions.

\section{Acknowledgements}
This research was supported by the National Science Foundation (grant number CHE-1453204) and used the Extreme Science and Engineering Discovery Environment (XSEDE), which is supported by the National Science Foundation (grant number ACI-1053575, allocation TG-CHE110009). AWG acknowledges support by the National Science Foundation (grant CHE-1416571). GRM acknowledges the Department of Education for support through the GAANN fellowship program. MAM was supported through the Predictive Theory and Modeling for Materials and Chemical Science program by the Office of Basic Energy Sciences (BES), Department of Energy (DOE) and under the auspices of the US DOE by LLNL under Contract DE-AC52-07NA27344. We thank Prof.~Rich Saykally for helpful discussions about the origin of the low frequency portion of the IR spectrum of liquid water.


%

\end{document}